\documentclass{article}

\usepackage{arxiv}

\usepackage[utf8]{inputenc} 
\usepackage[T1]{fontenc}    
\usepackage{hyperref}       
\usepackage{url}            
\usepackage{booktabs}       
\usepackage{amsfonts}       
\usepackage{nicefrac}       
\usepackage{amsmath,amsfonts,amssymb,algorithm,algorithmic}
\usepackage{microtype}      
\usepackage{lipsum}		
\usepackage{graphicx}
\graphicspath{ {./general_figs/} }
\graphicspath{ {./results/} }
\usepackage{amsmath}
\usepackage{natbib}
\usepackage{caption}
\usepackage{subcaption}
\usepackage{doi}
\usepackage{tikz}
\usetikzlibrary{positioning}
\usepackage{amsmath}
\DeclareMathOperator*{\argmax}{arg\,max}
\usepackage{setspace}
\usepackage{lineno}

\title{Human-Algorithm Collaborative Bayesian Optimization for Engineering Systems}

 \author{ \href{https://orcid.org/0000-0000-0000-0000}{\includegraphics[scale=0.06]{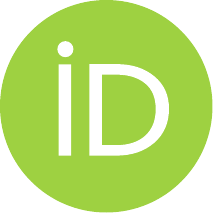}\hspace{1mm}Tom Savage}\\
	Department of Chemical Engineering\\
	Imperial College London\\
	\texttt{trs20@ic.ac.uk}
  \And 
  Ehecatl Antonio del Rio Chanona\\
	Department of Chemical Engineering\\
	Imperial College London\\
	\texttt{a.del-rio-chanona@imperial.ac.uk}}

\date{}

\renewcommand{\shorttitle}{Computers \& Chemical Engineering}

\begin{document}
\maketitle

\begin{abstract}
Bayesian optimization has been successfully applied throughout Chemical Engineering for the optimization of functions that are expensive-to-evaluate, or where gradients are not easily obtainable.
However, domain experts often possess valuable physical insights that are overlooked in fully automated decision-making approaches, necessitating the inclusion of human input. 
In this article we re-introduce the human back into the data-driven decision making loop by outlining an approach for collaborative Bayesian optimization.
Our methodology exploits the hypothesis that humans are more efficient at making discrete choices rather than continuous ones and enables experts to influence critical early decisions. 
We apply high-throughput (batch) Bayesian optimization alongside discrete decision theory to enable domain experts to influence the selection of experiments. 
At every iteration we apply a multi-objective approach that results in a set of alternate solutions that have both high utility and are reasonably distinct. 
The expert then selects the desired solution for evaluation from this set, allowing for the inclusion of expert knowledge and improving accountability, whilst maintaining the advantages of Bayesian optimization.
We demonstrate our approach across a number of applied and numerical case studies including bioprocess optimization and reactor geometry design, demonstrating that even in the case of an uninformed practitioner our algorithm recovers the regret of standard Bayesian optimization. 
Through the inclusion of continuous expert opinion, our approach enables faster convergence, and improved accountability for Bayesian optimization in engineering systems.

\end{abstract}


\section{Introduction}

Bayesian optimization has been successfully applied in complex domains where sampling is expensive and derivatives are often not available, such as those that involve experiments, simulation or propriety software. 
By removing the human from decision-making processes in favor of maximizing statistical quantities such as expected improvement, complex functions can be optimized efficiently \citep{tsay_bo,Zhang2024,Triantafyllou2023}. 
However, these engineering systems are often engaged with by domain experts such as engineers or chemists, and as such the behavior of the underlying function cannot be considered completely unknown a-priori \citep{Wang2022,Basford2024,Cho2024}. 
Therefore, there exists significant scope to leverage Bayesian optimization in optimizing expensive derivative-free problems, whilst enabling domain experts to inform the decision-making process, putting the human back into the loop. 

In previous work, \citet{Kanarik2023} demonstrated the potential for experts to enhance the convergence of Bayesian optimization when applied to a complex semi-conductor chip process, particularly during the initial exploration phase.
However, expert behavior, often consisting of changing one or two variables at a time, has been hypothesized to inhibit progress later on, where Bayesian optimization alone was deemed more efficient. 
\citet{Kanarik2023} highlights the difficulties in fusing expert and computational knowledge with respect to cultural challenges such as trust and transparency, and acknowledges the important role humans will play alongside computational tools.
Recent attempts have been made to integrate expert knowledge algorithmically within Bayesian optimization \citep{Liu2022,hvarfner2022pibo,hypbo}. 
\citet{hvarfner2022pibo} and \citet{Ramachandran2020} weight the acquisition function based on a user-specified prior in an effort to guide convergence towards expert-specified regions. 
In the case of a poor prior the approach has been found to be over-sensitive \citep{Liu2022}.
\citet{Liu2022} modified the approach by diminishing the weight of the expert-specified prior in the optimization, favoring it throughout the exploration-phase. 
By guiding the optimization towards the general `area' of human intuition, methodologies that weight the acquisition function ensure the expert has no guaranteed say in which solutions are evaluated throughout the optimization. 
Moreover, these approaches are considered static with expert-information only derived before the optimization.
Whilst updating the prior throughout optimization is an option to include human-feedback, this may be time-consuming for a practitioner and previous issues remain. Additionally, it is always possible to include a (simplified) model as the mean function to a Gaussian process, acting as a way to insert domain knowledge within the Bayesian optimization framework. 
\citet{BOMuse} presents an alternative approach, where an expert has continuous input throughout the optimization process. 
The expert, approximated by a linear Gaussian process, also selects a solution at a given stage, which is evaluated alongside the solution from a standard acquisition function.
By approximating human behavior, the authors present sub-linear regret bounds, faster than either the human-model or standalone Bayesian optimization. 
\citet{av2022human} demonstrates an approach where alternative solutions are proposed by both the expert and the acquisition function, and the information is combined to inform an evaluated experiment, in which the expert has the final say. 
Our work contributes broadly to the field of human-AI collaboration, in the same body of literature. \citet{reverberi2022experimental,hypbo} provides evidence that humans have the potential to internally condition information throughout complex medicine decision-making processes, to enhance AI systems, whilst \citet{Wang2019} provides an overview into the field of Human-AI collaboration within data-science. 

In this work, we attempt to remedy existing issues of including human-input, accountability, and ease-of-use by a methodology for human-in-the-loop Bayesian optimization that enables continuous expert input, whilst remaining practical for an expert to perform.
Importantly, our approach enables an expert to maintain influence over decision-making with minimal effort through a discrete selection selection step that may be performed remotely, quickly, and agnostic to platform (i.e. can be communicated efficiently online). 
Therefore removing a number of barriers of entry to applying human-in-the-loop methods in the real world such as the continuous time and effort an expert expected to contribute during the optimization process.

\begin{figure}[htb!]
    \centering
    \includegraphics[width=\textwidth]{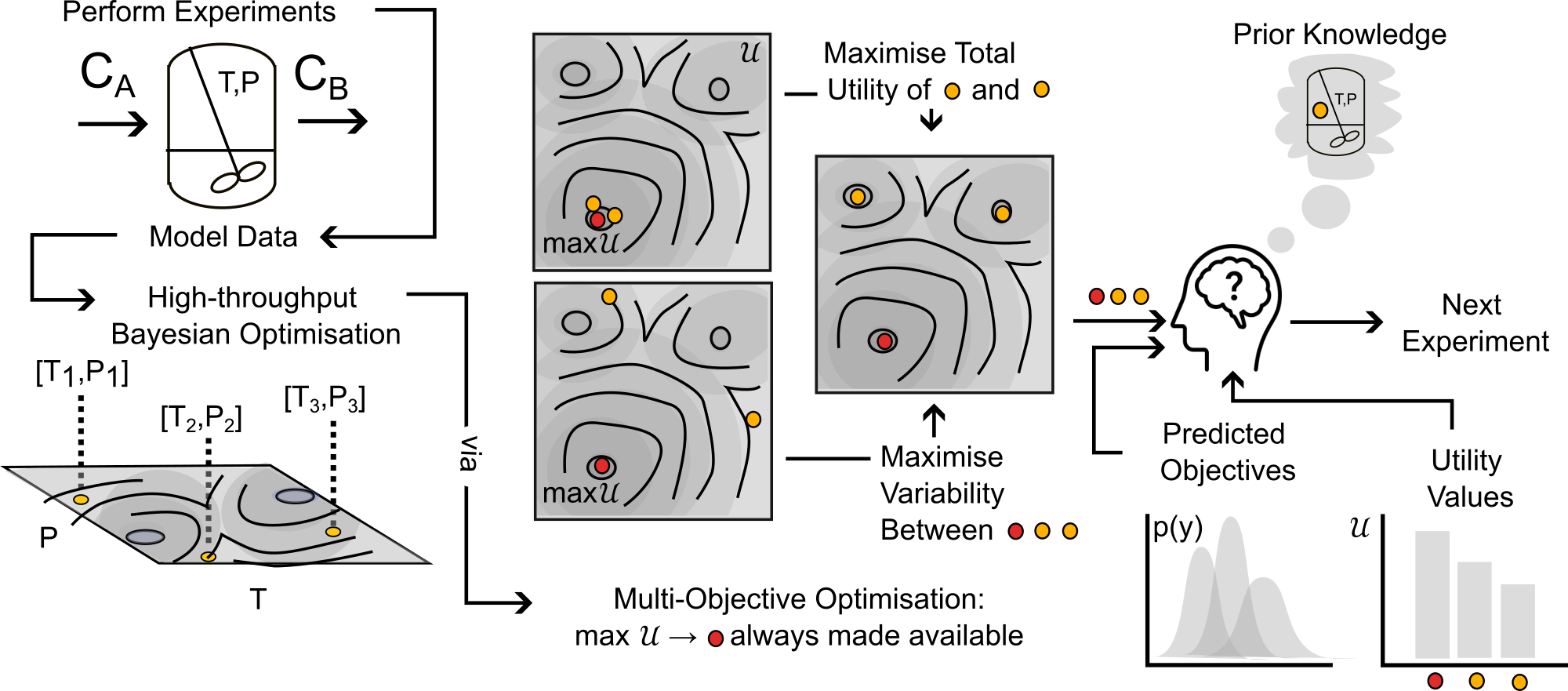}
    \caption{Methodology Overview.}
    \label{fig:overview}
\end{figure}

We apply high-throughput (batch) Bayesian optimization \citep{Gonzlez2023} alongside discrete decision theory to enable domain experts to influence the selection of optimal experiments. 
Our methodology exploits the hypothesis that humans are more efficient at making discrete choices rather than continuous ones, we also enable experts to influence important early decisions. 
At each iteration we solve a high-throughput (batch) multi-objective Bayesian optimization problem, optimizing the trade off between the sum of the alternative utility function values and the determinant of their covariance matrix, equivalent to their total variability. 
By taking the solution at the knee point of the Pareto front, we return a set of alternative solutions that have both high utility values and are reasonably distinct. 
Subsequently, we provide these alternatives to a domain expert to select their desired evaluation in the knowledge that any one choice presents information gain about the optimal solution, whilst ensuring the choices are distinct to the human expert. 
Alternative solution information such as utility function value, predicted output distribution, and visualizations are provided to the expert as prior knowledge. 
The decision-maker then effectively performs discrete Bayesian reasoning, internally conditioning their provided information with their own expertise and knowledge of the solutions provided (acting equivalently to a prior)\citep{Oaksford2009}. 
In addition, our approach enables improved interpretability in higher dimensions, as the decision-maker has the final say in what is evaluated. 
This approach works with any utility function and NSGA-II \citep{Deb2002} is applied for multi-objective optimization, efficiently handling the non-convex utility-space.  
Figure \ref{fig:overview} demonstrates our methodology. 

We benchmark our methodology on a number of numerical and applied case studies, and across different human behavioral policies including complete expertise, no knowledge, and an adversarial or misaligned practitioner. 
In doing so we provide insights into when expert-assisted Bayesian optimization approaches may be beneficial with respect to problem dimensionality, noise level, and function properties. 

\subsection{Contributions}

This work presents a novel methodology for human-in-the-loop Bayesian optimization that enables continuous expert input while remaining practical and easy to use. Our key contributions are as follows:
\begin{enumerate}
\item A collaborative approach exploiting humans' efficiency in making discrete choices.
\item Multi-objective high-throughput Bayesian optimization to balance utility and diversity.
\item Benchmarking insights for expert-assisted Bayesian optimization across various scenarios, demonstrating improved performance across a range of expert abilities.
\item Improved expert knowledge integration, accountability, and interpretability in decision-making.
\item Contribution to human-AI collaboration by removing barriers in real-world applications, demonstrated through case studies involving human experts in bioprocess optimization and reactor geometry design.
\end{enumerate}

This work represents a significant step towards the inclusion of domain experts within Bayesian optimization, linking discrete decision-making with state-of-the-art computational methodologies for the optimization of engineering systems.

\section{Background}

\subsection{Gaussian Processes}

A Gaussian process is an infinite-dimensional generalization of a multivariate Gaussian distribution, specified by a mean function $m(\cdot)$, and a positive semi-definite kernel (or covariance) function $k(\cdot,\cdot)$. 

\begin{align*}
\mathcal{GP} = \mathcal{N}(m(\cdot),k(\cdot,\cdot))
\end{align*}

A sample from a Gaussian process corresponds to an individual function. 
By conditioning the value of a Gaussian process on data at $n$ specific locations $\mathbf{X} \in\mathbb{R}^{d\times n}$, a Gaussian process of posterior functions can be derived.
Importantly, evaluating a Gaussian process at an input $\mathbf{x}\in\mathbb{R}^d$ will result in a one-dimensional Gaussian predictive distribution. 
This property enables the modeling of an underlying dataset probabilistically. 
For a given dataset of observations $\mathcal{D}:=\{(\mathbf{x}_i,y_i)\}_{i=0}^n$, the mean and variance of a Gaussian process posterior at location $\mathbf{x}$ are 

\begin{align*}
    \mu_{\mathcal{D}}(\mathbf{x}) = m(\mathbf{x}) + K(\mathbf{x},\mathbf{X})(\Sigma+\mathbf{N})^{-1}(\mathbf{y}-m(\mathbf{X}));\\
    \sigma_{\mathcal{D}}(\mathbf{x}) = K(\mathbf{x},\mathbf{x}) - k(\mathbf{x},\mathbf{X})(\Sigma+\mathbf{N})^{-1}K(\mathbf{X},\mathbf{x})
\end{align*}

where $k(\mathbf{x},\mathbf{X})$ is the covariance matrix between the input at $x$ and all other observations, and $\mathbf{N}$ is learnable matrix characterizing the noise of observations, often assumed to be the identity matrix multiplied by a scalar, i.e. $\sigma^2 \mathbf{I}$, characterizing a constant noise across all observations. 
Often the mean function $m(\mathbf{x})$, is assumed to be zero or a constant value which is learned, as to not induce any prior bias into the form of the function.

\subsection{Bayesian Optimization}

Bayesian optimization aims to solve the following problem
\begin{align}\label{bo_simple}
    \mathbf{x}^* \in \argmax_{\mathbf{x}\in\mathcal{X}} f(\mathbf{x})
\end{align}
where either the function $f$ is unknown or its form is not useful, gradients are not efficiently accessible, or function evaluations are expensive.
By modeling observations of $f$ using a Gaussian process, a distribution of the underlying function can be derived in the form of the posterior.
optimization under this uncertain model is then performed by evaluating a policy, specified by a utility function. 
By formulating the selection of the next evaluation of $f$ as an uncertain decision making problem (given the Gaussian process provides a probabilistic model), policies emerge that can be seen to balance exploration and exploitation of the search space.

\subsubsection{Expected Improvement}

In order to solve Eq. \ref{bo_simple}, a natural criteria is to apply distributional predictions to maximize the expected one-step-ahead improvement in the value of $f$.
Under the assumption that observations of $f$ are noiseless, this has the following analytical form and is referred to as the expected improvement acquisition or utility function \citep{Garnett2023}: 
\begin{align}\label{EI}
    \mathcal{U}_{\text{EI}}(\mathbf{x}|\mathcal{D}) = (\mu_{\mathcal{D}}(\mathbf{x})-\phi^*) \Phi\left(\frac{\mu_{\mathcal{D}}(\mathbf{x})-\phi^*}{\sigma_{\mathcal{D}}(\mathbf{x})}\right) + \sigma_{\mathcal{D}} \phi\left(\frac{\mu_{\mathcal{D}}(\mathbf{x})-\phi^*}{\sigma_{\mathcal{D}}(\mathbf{x})}\right)
\end{align}
where $\phi^*$ is the incumbent current best function value, $\Phi$ is the standard normal cumulative distribution function and $\phi$ is the standard normal distribution function. 
Recently, \citet{logEI} proposed a modification to Eq. \ref{EI} applying logarithmic quantities that provide more favorable conditions for optimization, overcoming issues that can result in sub optimal performance such as vanishing gradients.

When observations contain underlying measurement noise 
the expected improvement criteria does not have a concrete analytical expression. 
\citet{Garnett2023} outlines an approach which relies on the calculation of an upper envelope of a set of 1D lines at each evaluation, as well as a number of competing approximations \citep{hletham}. 

\subsubsection{Noisy Expected Improvement}
We outline the approach of \citet{hletham} that we will apply as a utility function in the case of noisy experimental data. 
Firstly, a Gaussian process is trained on the experimental data under the assumption that the measurements are noisy.
This Gaussian process is sampled at each input location, and the resulting data is treated as if it has no measurement noise. 
A Gaussian process is trained on this synthetic data under the assumption it is noiseless and the standard expected improvement function is formulated. 
A noisy approximation to expected improvement is to calculate an approximate expectation over Gaussian processes, each with a different set of synthetic data. 
Mathematically this approximation to noisy expected improvement is as follows \citep{Garnett2023}:

\begin{align}\label{letham_noise}
\mathcal{U}_{\text{EI-NOISY}}(\mathbf{x}|\mathcal{D}) \approx \int \mathcal{U}_{\text{EI}} (\mathbf{x};\mathbf{X},\mathbf{\phi}) \;p(\phi|\mathbf{X},\mathcal{D}) \;\text{d}\phi 
\end{align}

where $\phi$ represents the set of synthetic experimental outputs corresponding to input locations $\mathbf{X}$, that are treated as noiseless observations.

Whilst the approach demonstrated by \citet{hletham} has been deemed overly explorative, as the value of $\mathcal{U}_{\text{EI-NOISY}}(\mathbf{x}|\mathcal{D})$ drops to 0 in previously sampled locations, we find the approach easy to integrate with existing methodologies that apply automatic differentiation resulting in efficient optimization of the acquisition function, and fast to evaluate due to the ability to parallelize the evaluation of Eq. \ref{letham_noise} across different noise-less GPs. 
The ability to efficiently optimize an acquisition function has recently been highlighted as an important consideration in Bayesian optimization \citep{logEI}.
\citet{Picheny2013} provides an overview into a number of alternative noisy Bayesian optimization approaches and a discussion into their merits.

\section{Methodology}

In this section, we outline our approach for human-in-the-loop Bayesian optimization. 
Our methodology consists of 3 steps.
Firstly we outline an expert-informed initial design of experiments methodology.
Secondly we demonstrate a high-throughput (batch) multi-objective approach for Bayesian optimization in order to provide a set of alternative solutions. 
Finally we outline the human decision making procedure involved.

We first begin with a description of an optional step, wherein an expert provides a number of specific solutions that they wish to be evaluated. 
By performing this step before optimization begins, and not during optimization, we enable the expert to consider continuous solutions without prohibitive time constraints that may be present during optimization, as well as ensure that specific solutions are evaluated.
By including this step we incorporate benefits discussed in \citet{Kanarik2023} and \citet{reverberi2022experimental}.
We do so by applying an augmented initial design of experiments, enabling the number of specific solutions to be as large or small as the expert deems appropriate relative to the total number of space-filling points. 

\subsection{Expert-augmented Design of Experiments}

We first define an initial number of $m$ potential $d$-dimensional solutions as specified by an expert as $\mathbf{X}_{\text{expert}} \in \mathbb{R}^{m\times d}$. 
Subsequently we present the $t-m$ remaining solutions as $\mathbf{X}_{\text{design}}\in\mathbb{R}^{(t-m) \times d}$, where $t$ is the total number of initial solutions required. 
If $m \geq t$ then there is no need to proceed with the following step and the final design of experiments becomes $\mathbf{X}_{\text{expert}}$.
In order to distribute the remaining initial solutions throughout the design space we solve the following optimization problem: 

\begin{align}\label{dist}
   \min_{\mathbf{X}_{\text{design}}} -|K_{\mathbf{X}_\text{aug}}|
\end{align}
where $\mathbf{X}_\text{aug}$ represents the concatenation of $\mathbf{X}_{\text{design}}$ and $\mathbf{X}_{\text{expert}}$, and $K_{\mathbf{X}_{\text{aug}}} = [k(\mathbf{X}_{\text{aug},i},\mathbf{X}_{\text{aug},j})]^p_{i,j=1}$ is the covariance matrix between all solutions within the augmented set using a given kernel function (i.e. distance based).
Equation \ref{dist} reflects the principle of maximizing the determinant of the covariance matrix, equivalent to a 'volume of information' spanned by all solutions. 
By optimizing over the remaining solutions, we ensure these are evenly distributed throughout the design space, taking into account the fixed locations within $\mathbf{X}_{\text{expert}}$. 

\begin{figure}[htb!]
     \centering
     \includegraphics[width=\textwidth]{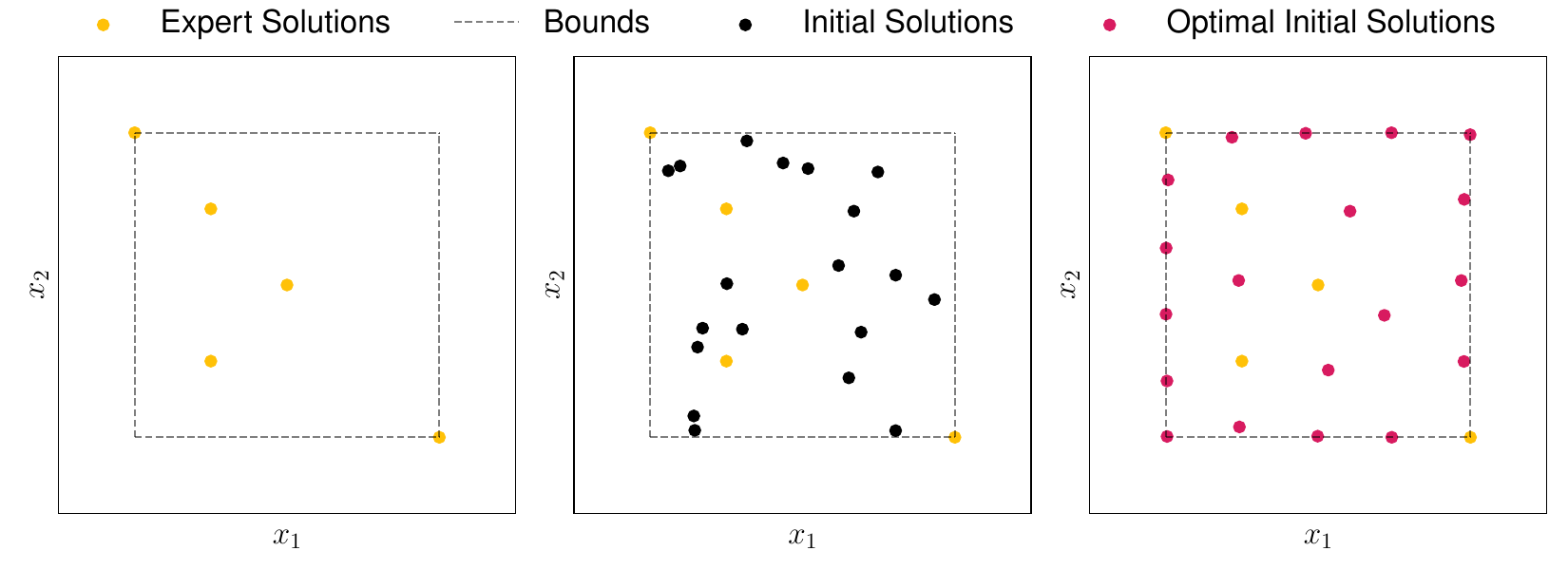}
     \caption{The expert-augmented design of experiments procedure outlined using a 2 dimensional example. \textbf{Left:} the expert defined set of solutions. \textbf{Center:} the initial random set of solutions defined as $\mathbf{X}_{\text{design}}$. \textbf{Right:} the optimal set of solutions comprising of both $\mathbf{X}_{\text{design}}$ alongside $\mathbf{X}_{\text{expert}}$, resulting in the final design of experiments.}
     \label{lhs_fig}
 \end{figure}

Figure \ref{lhs_fig} demonstrates our approach for expert-informed generation of an initial set of solutions for use as a dataset in Bayesian optimization. 

We now turn our attention to the Bayesian optimization loop itself, beginning with a focus on how to define alternative sets of solutions to provide to an expert at each iteration.

\subsection{Defining Alternate Solutions}

To begin our approach, at each iteration we first maximize a given utility function $\mathcal{U}$ for a given dataset $\mathcal{D}_t:= \{(\mathbf{x}_i,y_i)\}_{i=1}^t$:
\begin{align}\label{standard_bo}
   \mathbf{x}^* = \argmax_{x\in\mathcal{X}\subseteq\mathbb{R}^d} \; \mathcal{U}(x),
\end{align}
resulting in the optimal next evaluation, $\mathbf{x}^*$, from a standard utility sense (for example expected-improvement).
This step remains the same as in standard Bayesian optimization, where this solution would be evaluated, added to the dataset, and the procedure repeated. 

In our approach, the goal is to present a number of alternatives to the expert from which they may choose their desired solution. 
By considering utility values, and differences between these solutions, our goal is to ensure that all have large expected improvement values, whilst remaining distinct within the design space. 
Let $p$ be the number of alternate solutions provided to the expert and construct the decision variable matrix $\mathbf{X} \in \mathbb{R}^{(p-1)\times d}$ by concatenating $p-1$ alternate solutions $\mathbf{X} := [\mathbf{x}_1,\dots,\mathbf{x}_{p-1}]$.
We then define the high-throughput (batch) utility function $\hat{\mathcal{U}}$ which is specified as the sum of the individual utilities of alternate solutions within $\mathbf{X}$
\begin{align}
    \hat{\mathcal{U}}(\mathbf{X}) = \sum_{i=0}^{p-1} \mathcal{U}(\mathbf{X}_i).
\end{align}
Similarly, we introduce $\hat{\mathcal{S}}$ as a measure for capturing the variability among both the optimal and alternative solutions.
Specifically, let $\hat{\mathcal{S}}$ be the determinant of the covariance matrix $K_{\mathbf{X}_{\text{aug}}}$ for the augmented set $ \mathbf{X}_{\text{aug}}= \mathbf{X} \cup \mathbf{x}^*$:
\begin{align*}
 \hat{\mathcal{S}}(\mathbf{X},\mathbf{x}^*) &= |K_{\mathbf{X_{\text{aug}}}}| \\
 K_{\mathbf{X}_{\text{aug}}} &= [k(\mathbf{X}_{\text{aug},i},\mathbf{X}_{\text{aug},j})]^p_{i,j=1}
\end{align*}
$\hat{\mathcal{S}}$ quantifies the `volume of information' spanned by the alternative solutions $\mathbf{X}$ as well as the optimal solution $\mathbf{x}^*$.
Maximizing $\hat{\mathcal{U}}$ will result in all alternative solutions proposed being the same as $\mathbf{x}^*$, that is $[\mathbf{x}^*_1,\dots,\mathbf{x}^*_{p-1}]$.
Contrary to this, maximizing $\hat{\mathcal{S}}$ will result in a set of solutions that are maximally-spaced both with respect to other alternatives, but also to $\mathbf{x}^*$.
At iteration $t$, we then solve the following multi-objective optimization problem:
\begin{align}\label{multi-objective}
    [\mathbf{X}^*_{f_1},\dots,\mathbf{X}^*_{f_2}] = \max_{\mathbf{X}} \; \left(\hat{\mathcal{U}}(\mathbf{X};\mathcal{D}_t),\hat{\mathcal{S}}(\mathbf{X},\mathbf{x}^*)\right),
\end{align}
resulting in a set of solutions along the Pareto front of both objectives $f_1$ and $f_2$. 
From this we define $\mathbf{X}^*_{k}$ as the solution at knee-point of the Pareto front. 
The $p-1$ individual solutions contained within $\mathbf{X}^*_k$ optimally trade off the sum of their utility values, with their variability. 
This ensures that when provided to an expert, alongside $\mathbf{x}^*$, any individual solution will have high expected information gain, and the solutions themselves will be distinct enough to ensure the expert is not made to make an effective gradient calculation (i.e. solutions are very close and practically indistinguishable).   
The practitioner is then made to choose a solution to evaluate from this set of alternatives. 
To do so, they are provided with information such as the utility value of each solution, expected output distributions (obtained from the Gaussian process), and information regarding previous solutions that they may wish to draw upon. 
In doing so, the practitioner effectively performs an internal discrete Bayesian reasoning, conditioning previous prior information and expert opinion with the mathematical quantities provided to make an informed decision.

Algorithm \ref{alg:Human-Informed_BO} illustrates the methodology. Our approach allows for an expert to include a number of continuous solutions within the initial dataset $\mathcal{D}$, which design of experiments can be performed around.
Additionally, the algorithm may be used alongside other approaches which allow for an expert to specify a distinct solution to be evaluated at every iteration.

\begin{algorithm}
\caption{Expert-Guided Bayesian Optimization}
\label{alg:Human-Informed_BO}
\begin{algorithmic}[1]
    \STATE \textbf{Initialize:} Objective function \(f\), Initial data \(\mathcal{D}\), Domain \(\mathcal{X}\) 
    \STATE Alternative choices \(p\), Utility function \(\hat{\mathcal{U}}\), Variability function \(\hat{\mathcal{S}}\), Termination Criteria \\
    \vspace{2mm} 

    \WHILE{Termination Criteria is False}
        \STATE \(\mathbf{x}^* \leftarrow \argmax_{x\in\mathcal{X}} \mathcal{U}(x)\) \texttt{ // Standard Bayesian optimization Step} \\
        \vspace{2mm} 

        \STATE \texttt{\textbf{// Multi-Objective High-Throughput Optimization}} 
        \STATE \([\mathbf{X}^*_1,\dots,\mathbf{X}^*_m] \leftarrow \max_{\mathbf{X}} (\hat{\mathcal{U}}(\mathbf{X};\mathcal{D}), \hat{\mathcal{S}}(\mathbf{X},\mathbf{x}^*))\) \\
        \vspace{2mm} 

        \STATE \(\mathbf{X}^*_k \leftarrow \text{knee}([\mathbf{X}^*_1,\dots,\mathbf{X}^*_m])\) \texttt{ // Select Knee-solution} \\
        \vspace{2mm} 

        \STATE \texttt{\textbf{// Expert selects from alternatives and standard optimal}} 
        \STATE \(\mathbf{x}_{\text{eval}} \leftarrow \underset{\mathbf{x} \in \mathbf{X}^*_k\cup \mathbf{x}^*}{\text{expert}} \mathbf{x}\) \\
        \vspace{2mm} 

        \STATE \(\mathcal{D} \leftarrow \{(\mathbf{x}_{\text{eval}}, f(\mathbf{x}_{\text{eval}}))\} \cup \mathcal{D}\) \texttt{ // Update dataset} \\
    \ENDWHILE
\end{algorithmic}
\end{algorithm}

\subsection{Algorithm Characteristics}

\begin{figure}[htb!]
    \centering
        \begin{subfigure}[b]{\textwidth}
         \centering
        \includegraphics[width=\textwidth]{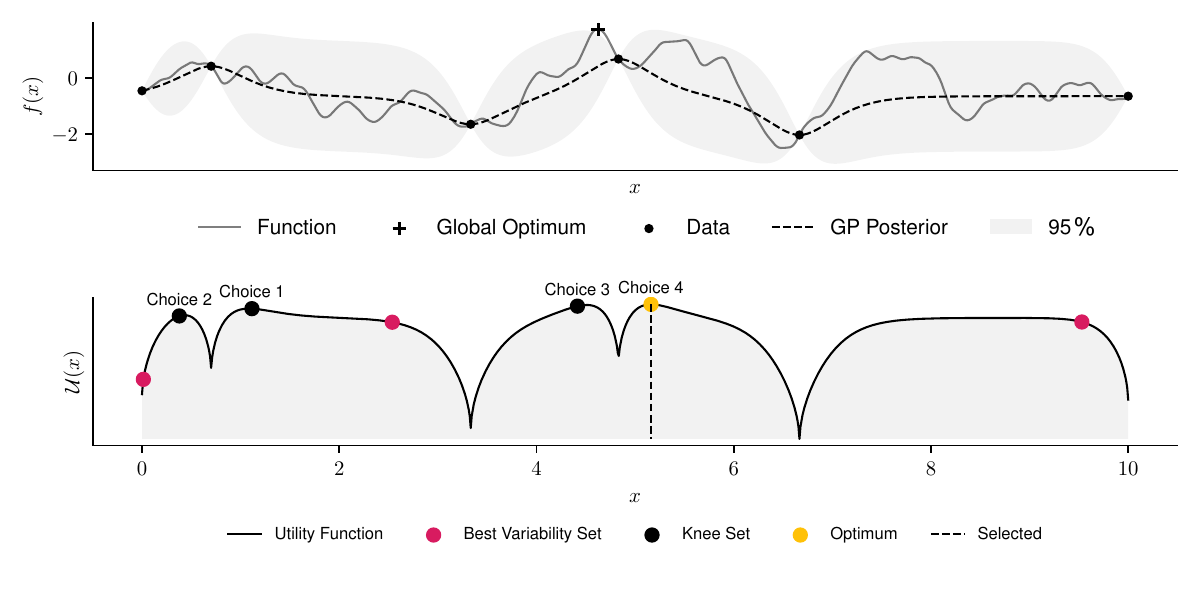}
         \caption{The objective function and utility function after 5 function evaluations. The 2 alternative solutions that maximize the solution distance can be seen in red, whilst the black solutions denote those contained within the knee-solution of the high-throughput multi-objective problem. The yellow optimal solution is included alongside these two alternatives to an expert. In this case choice 3 is selected randomly from the 3 alternatives.}
         \label{b_1_aq}
     \end{subfigure}
        \begin{subfigure}[b]{\textwidth}
         \centering
        \includegraphics[width=\textwidth]{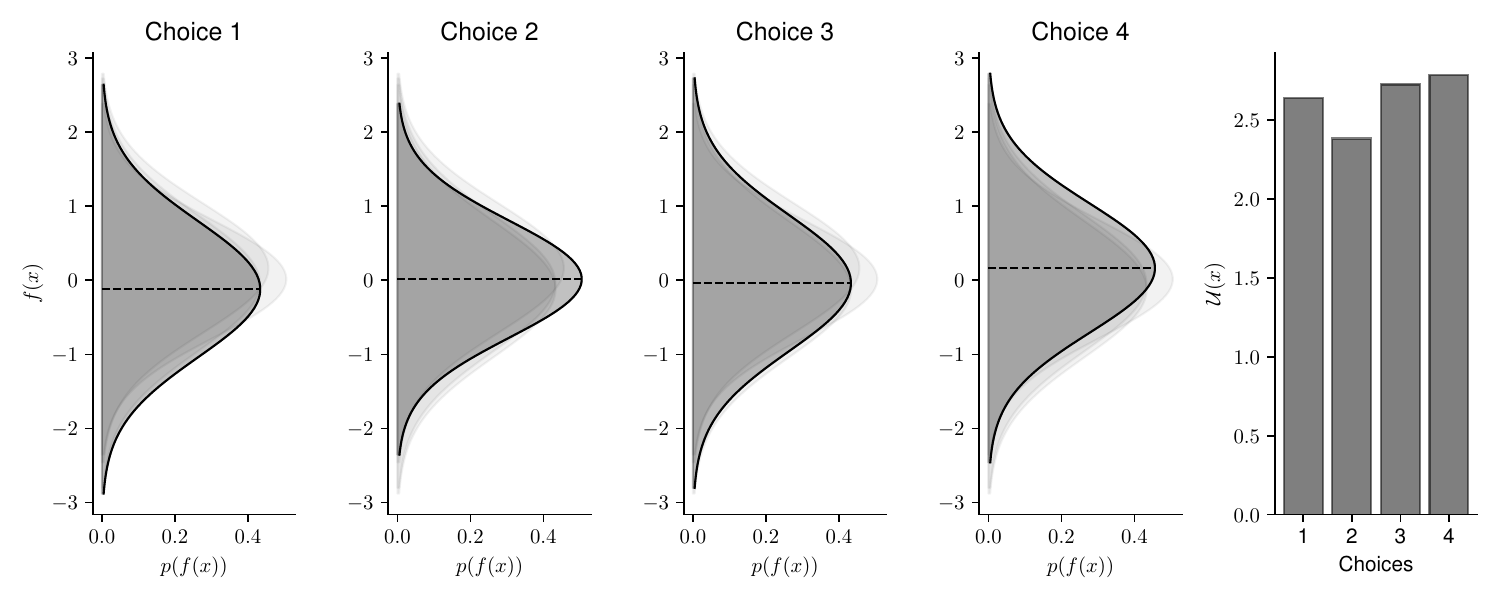}
         \caption{The information provided to the expert regarding the three alternative solutions. In this case choice 1 and choice 3 have relatively similar utility values and predicted output distribution than choice 4 (the optimal of the acquisition function). The expert is then allowed to distinguish between these similar solutions in a way the computer cannot through their prior domain knowledge. By conditioning their prior information on the given values, the expert is effectively performing internal Bayesian reasoning.}
         \label{b_1_choices}
     \end{subfigure}
     \caption{A standard iteration of our algorithm on a one-dimensional case-study.}
    \label{behaviour_1}
\end{figure}

Figure \ref{behaviour_1} demonstrates the intended behavior of our approach. We present a one-dimensional case study, optimizing a function obtained through sampling a Gaussian process prior, specified by a Mat\'ern 5/2 kernel with lengthscale $l=0.5$. In this case study we provide 3 alternatives to an expert, whose choice we select at random. 

To illustrate the importance of including the optimum of the utility function within the set of alternatives, consider a scenario where the utility function possesses a single optimal solution, representing a singular clearly desirable experiment. 
In such a case, the multi-objective approach, which focuses on maximizing both aggregate utility and variability could lead to suboptimal outcomes, as the optimal set of alternatives may attempt to straddle the single optimal experiment.
In this case, the solutions provided to the expert have ignored the clear solution in pursuit of increased variability.
By including the solution that maximizes the utility function within the set of alternatives and calculating the variance with respect to this, we mitigate against this behavior. 
In doing so we also allow the expert to simply fall back to standard Bayesian optimization by selecting the solution with the largest utility value. 

\subsection{Selecting from a Set of Solutions}

Having presented an approach to provide a set of promising and distinct alternative solutions to the expert, we draw our attention to how to present these to a practitioner. 
As previously mentioned, the expert effectively performs discrete Bayesian reasoning, conditioning the information provided with their prior beliefs of the underlying function. 
Whilst the utility function summarizes the predictive output distribution of a given solution, we choose to include the output distribution alongside utility values. 
This enables the practitioner to infer their own internal solution utility, improving interpretability. 
For example, the expert practitioner may have more interest in regions of high uncertainty, which, whilst potentially being sub-optimal from an expected improvement perspective, will provide them with the confidence to implement the resulting optimal solution. 

Solution information provided to the practitioner may be provided in a string-based format. 
However, for problems with a large number of variables this may become prohibitive. 
Therefore, we make the recommendation and extension that for applied high dimensional problems an image-based solution representation may be used.
For example in black-box bioprocess control, a visualization of the control profile as we illustrate in our case study.

\section{Case Studies}

We analyze the performance of our approach through a number of mathematical functions and real world chemical engineering case studies.
We use standard Bayesian optimization as a baseline for our comparisons.
The proposed methodology assumes human experts as part of the decision-making, therefore to characterize how our method will behave depending on the level of expertise of the human making decisions, we hypothesize a number of different human behaviors.
The `Expert' practitioner represents an ideal, where the best solution (that is the one with the highest true objective value) is always selected. 
Equivalently, to test the performance of our approach under the influence of a practitioner with misaligned knowledge, we present an `Adversarial' practitioner who consistently selects the solution with the lowest true objective value.
In addition, we present a probabilistic practitioner, who selects the solution with the best true objective value with some probability. 
In practice, the human will condition the information provided (e.g. objective function value, visuals of the possible solutions) with their prior beliefs (e.g. expert knowledge). 
In our approach this includes information regarding the expected distribution of the objective of each solution, as well as the Utility value of each solution. 
Therefore, we present a final behavior of a `Trusting' practitioner who selects the solution with the largest utility, equivalent to standard Bayesian optimization as this solution is obtained through standard single objective optimization. 
The human behaviors applied are summarized within Table \ref{behav}.

\begin{table}[h]
\centering
\caption{Human Behaviors Applied for Benchmarking}
\label{behav}
\renewcommand{\arraystretch}{1.5} 
\begin{tabular}{ll}
\hline
\textbf{Behavior Type} & \textbf{Description}                                                                                                                                       \\ \hline
Expert                                  & Selects the solution with the best true function value.                                                                                                                     \\
Adversarial                             & Selects the solution with the worst true function value.                                                                                                                    \\
Trusting                                & Selects the solution with the maximum utility value.                                                                                                                        \\
\(p(\text{Best})\)                      & \begin{tabular}[c]{@{}l@{}}Selects the solution with the best true function value with \\ probability \(p(\text{Best})\), otherwise selects a random solution.\end{tabular} \\ \hline
\end{tabular}
\end{table}

\subsection{Benchmark Functions}
We benchmark our approach on a number of well-known optimization functions. For each standard optimization benchmark function we evaluate the performance of different expert behaviors for its 2 dimensional, 3 dimensional, and 5 dimensional form.
The equations for these test functions can be located within the Appendix. 

In addition, we generate functions by sampling from a Gaussian process prior, enabling us to benchmark our approach in expectation across all functions of a given lengthscale. 
We generate random 2, 3 and 5-dimensional functions from priors with a lengthscale of 0.05 within the bounds 0 and 1. 
A visualization of lower-dimension sampled functions can be located in the Appendix.

For each test function, we perform Bayesian optimization across a number of noise levels, where the standard deviation of the noise represents 0\%, 5\% and 10\% of the range of each function (difference between the $\min f$ and $\max f$). 

We compare our approach for both UCB and EI utility functions. 
When $\sigma > 0$ (Gaussian distributed observation noise) the approximation presented in Eq. \ref{letham_noise} is applied with the expectation taken over either the UCB or EI acquisition function across 8 noiseless Gaussian processes, each trained with a sampled set of outputs from a Gaussian process trained with noise.
Gradient-based optimization of the acquisition function is performed using a multi-start of 256 initial solutions evaluated in parallel, using the L-BFGS-B solver \citep{zhu1997algorithm} and a low tolerance of 1E-10.
The multi-objective problem outlined in Eq. \ref{multi-objective} is solved using the NSGA-II algorithm \citep{Deb2002} with a population size of 50. 
Given these are test functions we opt not to take advantage of the augmented initial design of experiments step, instead initial solutions are generated using a static Latin-hypercube design-of-experiments with 8 solutions.
For each standard optimization test function and respective noise level, we run Bayesian optimization for 48 iterations a total of 32 times with different initial conditions, and take the average regret across all runs.
When optimizing the functions from a Gaussian process prior of a given lengthscale and dimension, we sample 50 unique functions and keep the initial conditions the same across all runs. 

\subsection{Pulsed-flow Coiled Tube Reactor Optimization}

In addition to benchmark problems, we highlight the effectiveness of our approach on two applied chemical engineering black-box optimization tasks.

The first task concerns the joint optimization of operating conditions and geometry of a pulsed-flow helical tube reactor. 
Operating conditions such as pulse frequency, and geometric parameters such as coil radius and pitch have previously been shown to have a significant impact on the plug-flow performance in coiled tube reactors at low Reynolds numbers. 
By optimizing these parameters in order optimize the shape of the residence time distribution, the efficiency and product quality of mesoscale flow processes can be greatly improved.
Each function evaluation consists of simulating a tracer experiment within OpenFOAM for a given reactor and operating conditions, resulting in an expensive (approximately 1 hour per simulation) and derivative-free function. 
More information into this specific case study can be located within \citep{Savage2023,Basha2023}. Table \ref{reac_tab} shows all optimization parameters and respective bounds. 

\begin{table}[htb!]
\centering
\caption{The parameters and respective bounds for the optimization of }
\renewcommand{\arraystretch}{1.5} 
\begin{tabular}{llllll}
\hline
 & \multicolumn{1}{c}{\textbf{a (m)}} & \multicolumn{1}{c}{\textbf{f (Hz)}} & \multicolumn{1}{c}{\textbf{Re}} & \multicolumn{1}{c}{\textbf{p (m)}} & \multicolumn{1}{c}{\textbf{r (m)}} \\ \hline
Lower Bound & 0.001 & 2 & 10 & 0.0075 & 0.005 \\
Upper Bound & 0.008 & 8 & 50 & 0.015 & 0.015 \\ \hline
\end{tabular}\label{reac_tab}
\end{table}

Where \textit{a} is the amplitude of pulsed-flow at the inlet of the reactor, \textit{f} is the pulsed-flow frequency, \textit{Re} is the Reynolds number, which in turn defines the velocity at the inlet of the reactor. Finally, \textit{p} and \textit{r} are the pitch and radius of the coil respectively.

In order to enable rapid benchmarking, 75 reactor configurations were distributed throughout the design space using a Latin hypercube design of experiments, and each evaluated in OpenFOAM. 
A Gaussian process was then trained using this data, and the resulting mean used as the ground-truth function. 

\begin{figure}[htb!]
    \centering
    \includegraphics[width=\textwidth]{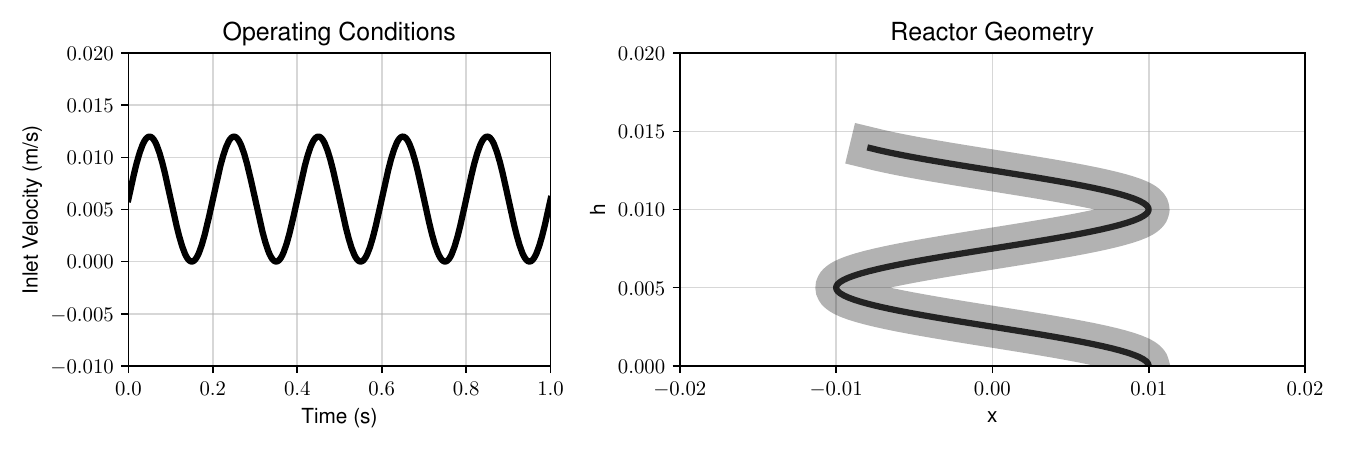}
    \caption{A reactor configuration visualized in order to aid the expert in making discrete choices. In doing so, the expert had a reduced need to interpret higher-dimensional decision vectors.}
    \label{reac_example}
\end{figure}

Whilst we cannot provide a result visualization for this case study due to the application of an approximation to the true underlying function, we include a solution visualization to assist the expert in making discrete choices. 
An example of this solution visualization is shown in Fig \ref{reac_example}.

\subsection{Black-box Bioprocess Optimization}


This task concerns the fed-batch optimization of a phycocyanin synthesised by cyanobacterium Arthrospira platensis via photo-production.
To evaluate the system, non-smooth bioprocess dynamics are integrated for a total of 200 simulated hours. 
Details of the dynamics can be located in the work by \citet{Petsagkourakis2020}.
For the purposes of this case study we assume that the dynamics of the reaction are not able to reliably be learned, due to the time involved to collect data and the small amount of data, necessitating a black-box approach.

Both nitrogen feed rate and light intensity serve as the optimization variables, and can be changed 8 equally spaced times throughout the time-horizon at values between 0-400 mgL\textsuperscript{-1} and 0-40 Lux respectively. 
The objective is to maximize the total end concentration of phycocyanin product.

This black-box bioprocess case-study represents an expensive high-dimensional experimental system with a nonlinear objective, where domain knowledge may be used alongside other contextual information to provide potential improvements.

\begin{figure}[htb!]
    \centering
    \includegraphics[width=\textwidth]{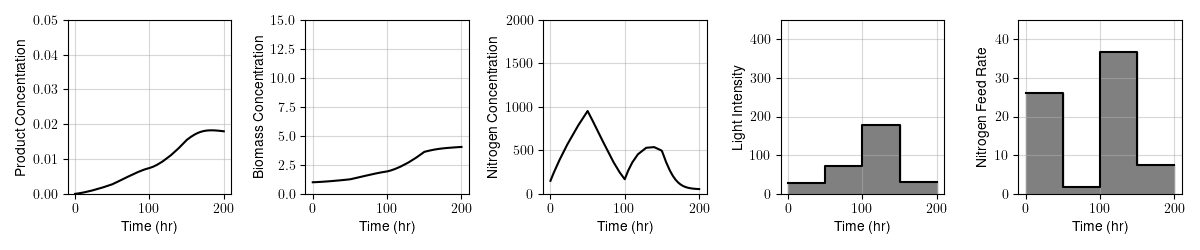}
    \caption{A reactor configuration visualized in order to aid the expert in making discrete choices. In doing so, the expert had a reduced need to interpret higher-dimensional decision vectors.}
    \label{bioproc_example}
\end{figure}

To aid the expert in making a discrete decision we provide a visualization of the proposed control scheme. 
In addition, as this is a simulated case study, we postulate that online measurements are available after a control scheme has been evaluated. 
This additional information, which is reasonably available in the engineering domain, may enable the expert to reason about the choices based on previous evaluations more easily.
Figure \ref{bioproc_example} demonstrates this result visualization made available for all evaluated control schemes.

\begin{figure}[htb!]
    \centering
    \includegraphics[width=0.45\textwidth]{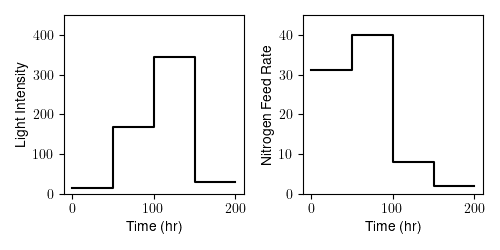}
    \includegraphics[width=0.45\textwidth]{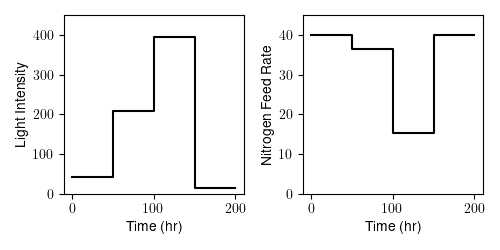}
    \includegraphics[width=0.45\textwidth]{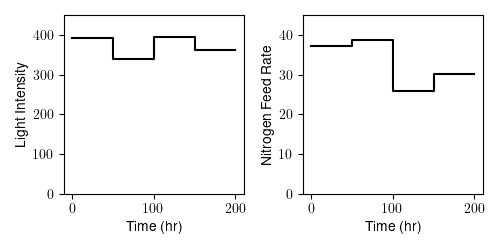}
    \includegraphics[width=0.45\textwidth]{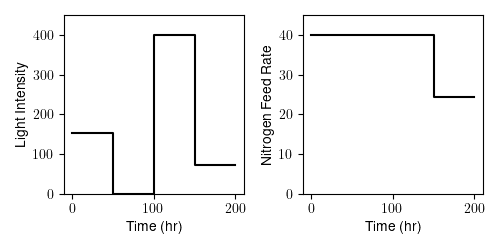}
    \caption{Four alternative control solutions, as made available to the expert in order to facilitate easier reasoning particularly in higher dimensions.}
    \label{cont_example}
\end{figure}

Figure \ref{cont_example} demonstrates an example of the information made available about the four alternative solutions, to the expert. 

The optimization variables and respective bounds are shown in Table \ref{bioprocvar}.

\begin{table}[htb!]
\centering
\caption{The parameters and respective bounds for the optimization of a fed batch bioprocess.}
\renewcommand{\arraystretch}{1.5} 
\begin{tabular}{lllllllll}
\hline
\multicolumn{1}{c}{\textbf{Bound}} & \multicolumn{1}{c}{\textbf{\begin{tabular}[c]{@{}c@{}}I\\ (0-50)\end{tabular}}} & \multicolumn{1}{c}{\textbf{\begin{tabular}[c]{@{}c@{}}I\\ (50-100)\end{tabular}}} & \multicolumn{1}{c}{\textbf{\begin{tabular}[c]{@{}c@{}}I\\ (100-150)\end{tabular}}} & \multicolumn{1}{c}{\textbf{\begin{tabular}[c]{@{}c@{}}I\\ (150-200)\end{tabular}}} & \multicolumn{1}{c}{\textbf{\begin{tabular}[c]{@{}c@{}}Fn\\ (0-50)\end{tabular}}} & \multicolumn{1}{c}{\textbf{\begin{tabular}[c]{@{}c@{}}Fn\\ (50-100)\end{tabular}}} & \multicolumn{1}{c}{\textbf{\begin{tabular}[c]{@{}c@{}}Fn\\ (100-150)\end{tabular}}} & \multicolumn{1}{c}{\textbf{\begin{tabular}[c]{@{}c@{}}Fn\\ (150-200)\end{tabular}}} \\ \hline
L & 0 & 0 & 0 & 0 & 0 & 0 & 0 & 0 \\
U & 400 & 400 & 400 & 400 & 40 & 40 & 40 & 40 \\ \hline
\end{tabular}\label{bioprocvar}
\end{table}

Where \textit{I} represents the constant value of light intensity between two fixed times denoted within the table, and \textit{F\textsubscript{n}} represents the respective inflow rate of nitrogen to the reactor. 

We note that that underlying dynamics are not made available to the humans performing the selection of the solutions, and only basic contextual information is provided that would reasonably be available to an expert.

\begin{figure}[htb!]
    \centering
    \includegraphics[width=\textwidth]{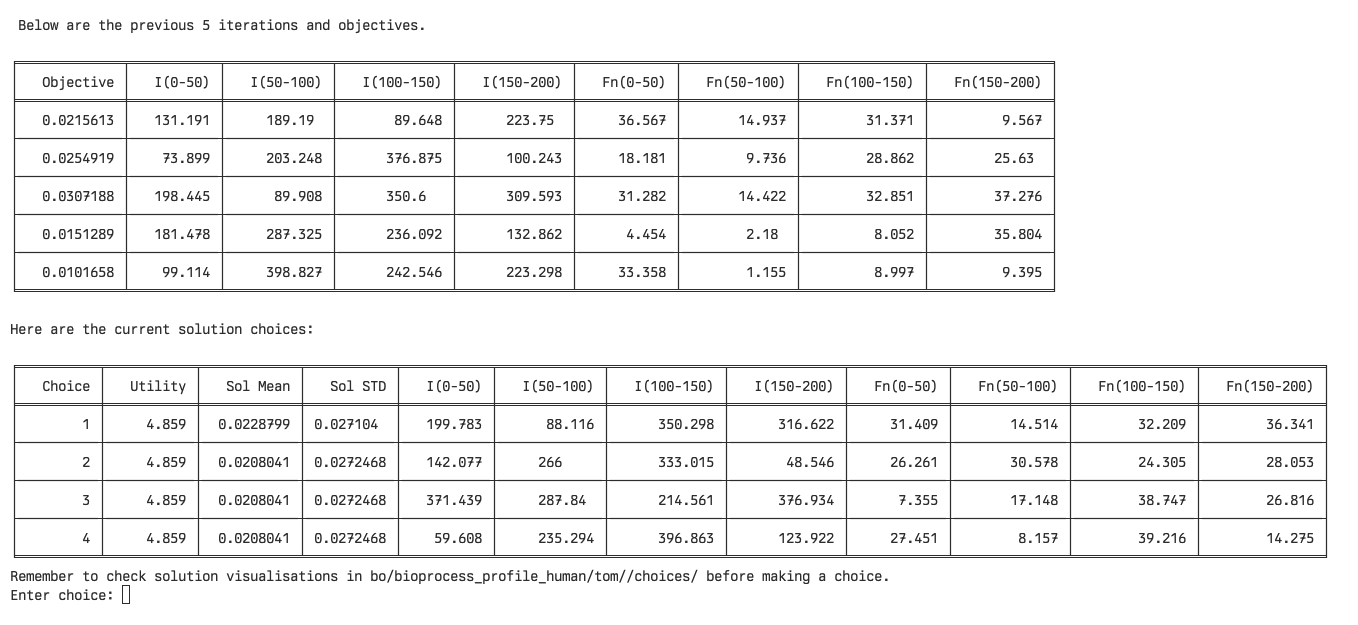}
    \caption{An instance of the textual information made available to the expert at each iteration.}
    \label{text}
\end{figure}

Figure \ref{text} demonstrates an instance of the textual information made available to the expert at each iteration, including information regarding the previous iterations and the choices available.

All code can be located within the following Github repository \url{https://github.com/OptiMaL-PSE-Lab/humbo/}. 

\subsection{Human-in-the-loop Experimental Protocol}

To evaluate the performance of our expert-guided Bayesian optimization approach with real human input, we conducted experiments with human participants on the fed-batch bioprocess optimization and reactor design case studies. 

All participants were chemical engineering postgraduate students with knowledge of optimization, design, and process control. 
Each participant was provided with an introduction to the problem, including the context, optimization variables, objective function, and any relevant background information that an expert in the field would typically have access to. 
Participants were then given the opportunity to select up to 8 initial solutions to initialize the experimental design. 
They were encouraged to apply their domain knowledge in making these selections.

Participants then engaged in the iterative Bayesian optimization process, consisting of: Reviewing information from previous iterations, including visualizations of evaluated control strategies and their corresponding objective values; Analyzing the current set of alternatives, which included predicted performance metrics (e.g., mean objective value) and visualizations of the proposed control strategies; Selecting the most promising solution from the alternatives based on their expertise and the provided information.

The selected solution was evaluated, and the results were added to the dataset for the next iteration.
The optimization process continued for 32 iterations.
Objective values and convergence metrics were recorded throughout the optimization process for each participant.
In addition, qualitative data such as comments and questions were recorded throughout.

For comparison, standard Bayesian optimization (without human input) was performed 32 times for each case study to generate a baseline for expected convergence.
The experiments were designed to mimic a realistic expert-guided optimization scenario while providing controlled conditions for comparing the performance of human-in-the-loop optimization against standard Bayesian optimization.

\section{Results \& Discussion}

We first present the effect of different problem factors, enabling us to conclude on the performance of our expert-guided collaborative approach across dimensionality, noise level, specific properties, and number of alternative solutions presented to the expert.

The adversarial or misaligned selection policy presents an upper bound on the rate of convergence, and the expert policy provides a lower bound on rate of convergence for non-myopic Bayesian optimization throughout the majority of results. 
Generally, increasing the ability of the expert, by increasing the probability of selecting the best choice results convergence between these upper and lower bounds. 
Standard Bayesian optimization which we deem the `trusting' policy (selects the solution with the largest utility value) also falls between these bounds.
By comparing this standard convergence with the stochastic expert policies we are able to make conclusions as to the potential of our expert-guided approach to improve convergence across a range of problem dimensionalities, noise levels, and types. 
Specifically, this enables us to conclude on when our approach is favorable over standard Bayesian optimization with respect to expert ability, problem class, and number of alternative solutions.

\subsection{Effect of Noise}

\begin{figure}[htb!]
    \centering
    \includegraphics[width=\textwidth]{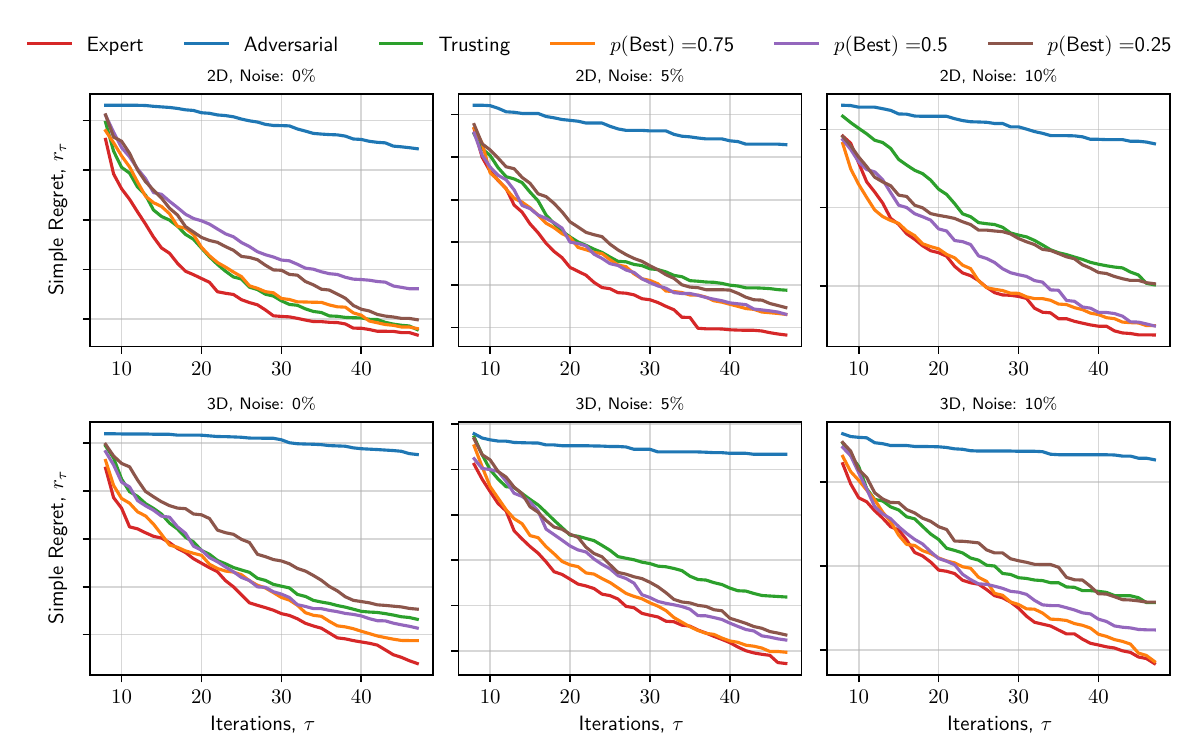}
    \caption{The effect of the level of observation noise on the difference in convergence for various expert behaviors across general 2D and 3D functions of a given length scale.}
    \label{noise}
\end{figure}

To investigate the effect of objective noise, Figure \ref{noise} demonstrates the results of our approach across general 2D and 3D objective functions sampled from a Gaussian process prior, with differing levels of noise.
We choose to display only the mean performance as opposed to including the standard deviation of regret to maintain the interpretability of the results.

Observing the convergence results for 2D functions, with no observational noise convergence is only faster than standard Bayesian optimization when the expert selects the best solution with 75\% accuracy.
At a 5\% observational noise level, the required accuracy to recover standard convergence drops to 25\% with 4 alternate choices, reflecting the noise level at which an expert selecting randomly recovers the standard level of convergence. 
This trend continues when increasing to 10\% observational noise, where an expert who selects randomly becomes more favorable than standard Bayesian optimization.

In higher-dimensional noisy problems, the quality of the expert required to recover the convergence of standard Bayesian optimization reduces across all noise levels. 
Performance across all noise levels across general 3D functions is similar, indicating that benefits are maintained in the presence of noise.

Based on these results we can conclude that our approach becomes more competitive for problems with greater observation noise, as the quality of expert required to improve on standard convergence becomes lower. 

In addition, we hypothesize that our methodology is favorable in noisy Bayesian optimization settings as it enables an expert to separate signal to noise internally, and to focus on prior `non-noisy' trends, reducing the overall number of iterations.

\subsection{Effect of Dimensionality}

\begin{figure}[htb!]
    \centering
    \includegraphics[width=\textwidth]{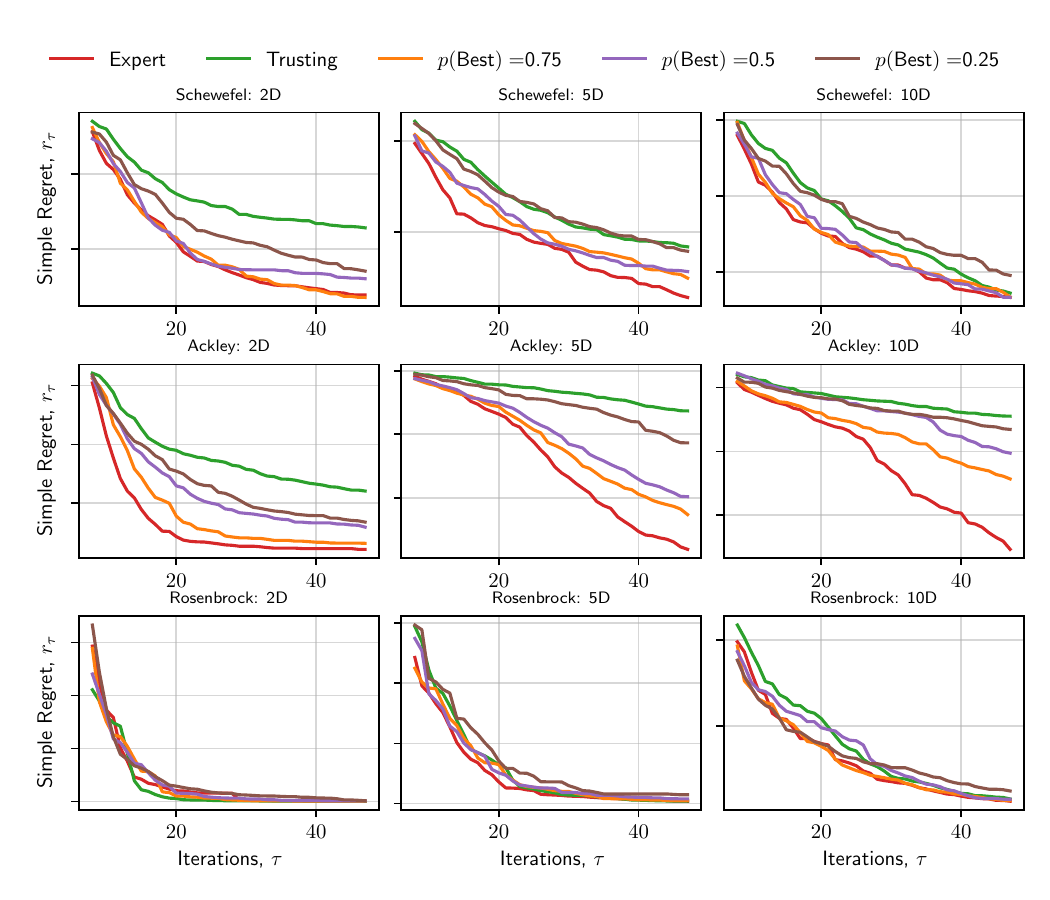}
    \caption{The effect of the dimensionality of the problem on the difference in convergence for various expert behaviors.}
    \label{dim}
\end{figure}

To demonstrate the effect of problem dimensionality on our approach, Figure \ref{dim} presents the convergence results with respect to 2, 5 and 10 dimensional functions with 2.5\% noise using the expected improvement acquisition function.
We have omitted the adversarial policy to improve interpretability.

General convergence behavior across all functions and increasing dimensionality is similar. 
The convergence of a random expert ($p=0.25$) matches or exceeds standard Bayesian optimization (trusting), indicating that in some problem settings the increased exploration that results from the random selection of a choice is beneficial.
As dimension increases the gap between the performance of an expert practitioner and standard Bayesian optimization diminishes. 
However, any expert that performs better than randomly still improves on the convergence of standard Bayesian optimization. 

This would indicate that the benefits of human-in-the-loop approaches diminish in higher dimensional cases, disregarding the ability to select specific solutions at the beginning of the process, something we omit for benchmarking purposes.

These results demonstrate that benefits of our approach are maintained in reasonably high dimensional and practical problem settings.

\subsection{Effect of Number of Alternative Solutions}

\begin{figure}[htb!]
    \centering
    \includegraphics[width=\textwidth]{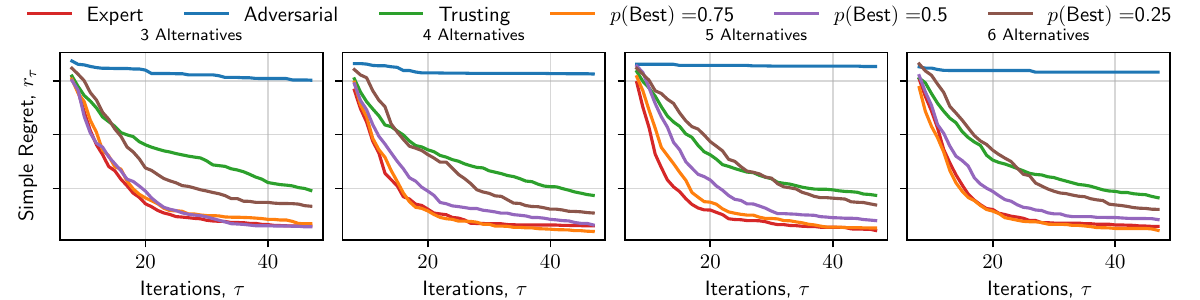}
    \caption{The effect of the number of alternative solutions made available on convergence for various expert behaviors.}
    \label{choices_results}
\end{figure}

We change the number of alternative solutions made available the hypothesized experts to investigate the performance of our approach.
We note that making more choices available has the potential to increase performance, as more solutions are made available to the expert, any one of which may be spotted as high performing. 
However, a human will be able to reason less effectively about the relative differences between a larger amount of solutions and may perform worse, indicating a trade-off in practice. 
However, when benchmarking computationally there is no consideration as to the increased difficult ability to reason about larger numbers of solutions due to the assumption of an underlying knowledge of the true function.
Figure \ref{choices_results} demonstrates the convergence results when increasing the alternate solutions from 3 to 6 for the two dimensional Ackley function.

Convergence generally becomes faster as a larger number of alternative solutions are made available at each iteration. 
Adversarial practitioners become slightly worse with more potential solutions, as there is more scope to make a poor selection. 
Whilst overall choice increases alongside the potential for improvement, it becomes more important to ensure that the expert is not misaligned. 

To summarize, a larger number of alternatives provides increased scope for convergence improvements.
However, at the same time it becomes more important to be confident in the quality of the expert, due to increased risk of misguided selections.

\subsection{Effect of Specific Function Properties}

\begin{figure}[htb!]
    \centering
    \includegraphics[width=\textwidth]{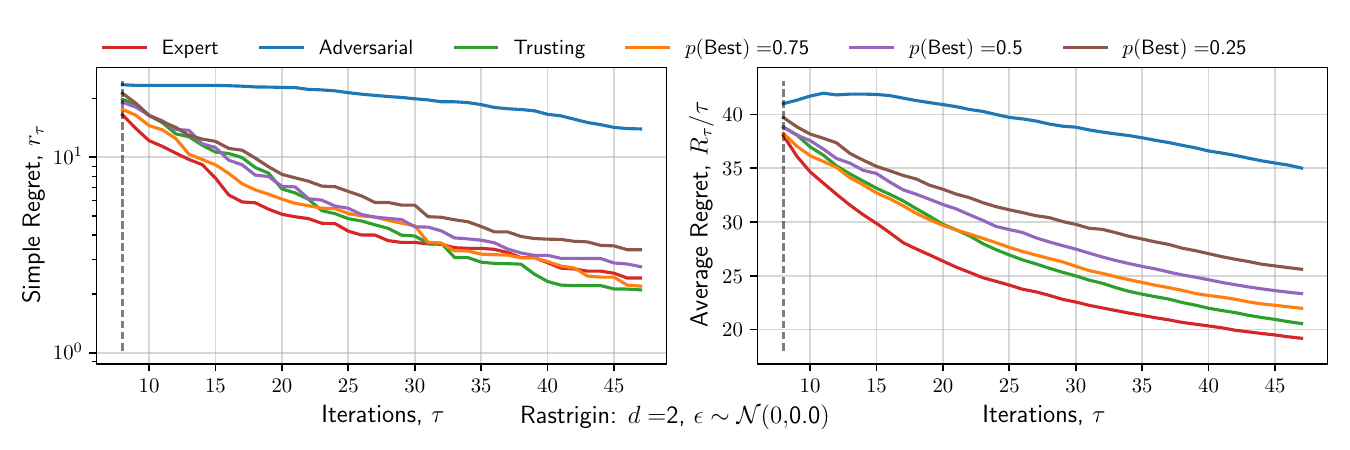}
    \caption{Convergence of expert-guided Bayesian optimization of the 2D Rastrigin function with no observation noise.}
    \label{rastrigin2}
\end{figure}

The Rastrigin function \citep{rastrigin1974systems} is highly multi-model, with a wide bowl-shape punctuated by individual peaks. 
Figure \ref{rastrigin2} shows convergence on the 2D Rastrigin function with no observation noise.

The convergence distribution for all policies are tight, with the exception of the hypothetical adversarial practitioner. 
Initially, expert policies have the same performance than standard Bayesian optimization, highlighting the ability for experts to influence Bayesian optimization in early stages. 
However, this benefit is negated in the later stages of optimization. 

\begin{figure}[htb!]
    \centering
    \includegraphics[width=\textwidth]{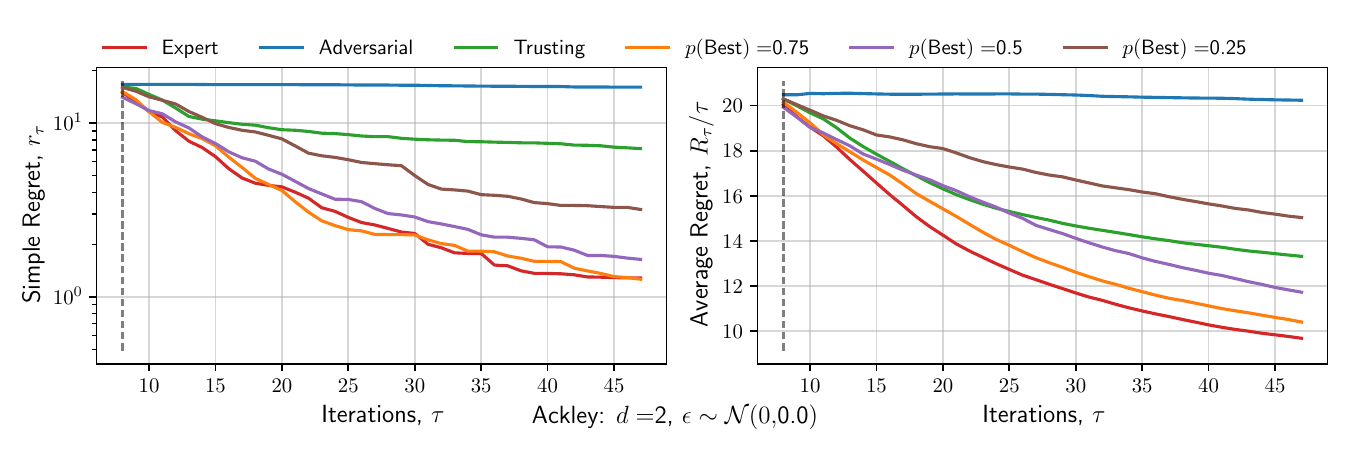}
    \caption{Convergence of expert-guided Bayesian optimization of the 2D Ackley function with no observation noise.}
    \label{ackley2}
\end{figure}

Whilst the Ackley function has similar multi-model characteristics to the Rastrigin function, the underlying form is less bowl shaped, resulting in the the value of the global optima being significantly different to the value of other local optima. 
Figure \ref{ackley2} shows convergence on the 2D Rastrigin function with no observation noise.

The distribution of convergence characteristics between policies is widened, and therefore the ability for the expert to provide a improvement is increased. 
In fact, selecting randomly from the 4 alternate solutions improves on standard Bayesian optimization, with policies that are better than random providing further improvements. 
We hypothesize that in this situation the underlying GP may have a higher length scale than is representative of the underlying function, enabling the expert to spot high performing solutions that are not predicted. 
Differences between these two functions, and the increased expert-guided exploration highlights the ability of our approach to identify solutions in situations where the value of the global optima is potentially significantly different to the value of other local optima.

\begin{figure}[htb!]
    \centering
    \includegraphics[width=\textwidth]{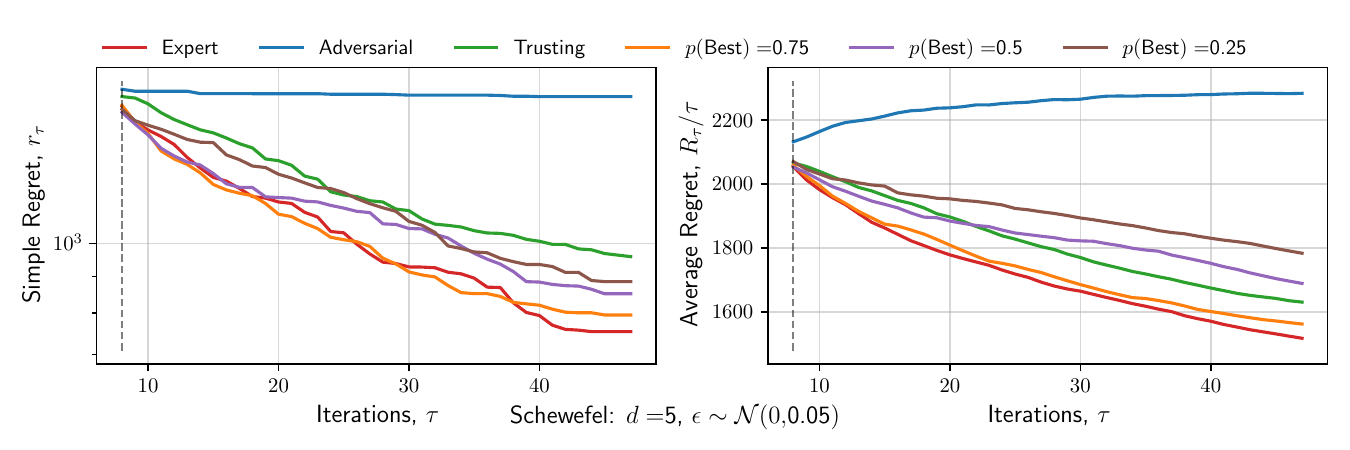}
    \caption{Convergence of expert-guided Bayesian optimization of the 5D Schewefel function with 5\% observation noise.}
    \label{sch}
\end{figure}

Finally, we highlight our approach on a more complex 5 dimensional multi-modal case with 5\% observation noise.
Figure \ref{sch} shows convergence on the 5D Schewefel function with 5\% observation noise.

The random expert with a 25\% chance of selecting the best solution out performs standard Bayesian optimization, indicating that in high-dimensional multi-model problems the additional exploration achieved through the selection of alternate solutions is favorable.

\subsection{Coiled Tube Reactor Design}
As well as stochastic expert policies which enable us to benchmark our approach with respect to expert ability, and problem class, we apply our approach on a black-box chemical engineering case study with real humans. 
These subjects are all chemical engineering postgraduate students, with knowledge of optimization, design, and broadly control.

\begin{figure}[htb!]
    \centering
    \includegraphics[width=\textwidth]{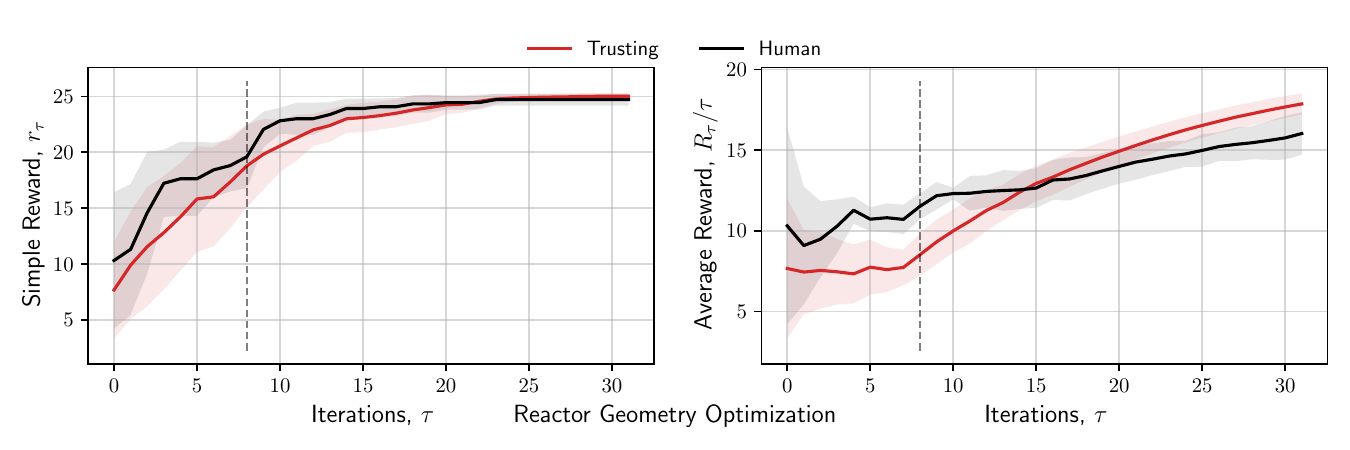}
    \caption{Simple and average reward, corresponding to the number of equivalent tanks-in-series for a given reactor geometry and operating conditions, presented alongside the reward obtained from standard Bayesian optimization. Filled sections represent one standard deviation and solid lines represent the mean reward.}
    \label{fig:reactor}
\end{figure}

Six participants were used to generate the human results, and standard Bayesian optimization was performed 32 times to generate the expected convergence metrics.
Figure \ref{fig:reactor} presents the convergence results for standard Bayesian optimization and our approach when simultaneously optimizing the geometry and operating conditions of a pulsed-flow coiled tube reactor. 

By applying collaborative Bayesian optimization we are able to locate reactor configurations with a larger equivalent tanks-in-series, maintaining an improvement over standard Bayesian optimization throughout the design process.
As the iterations progress the advantage diminishes, consistent with observations in literature.
Whilst average reward is more commonly applied in reinforcement learning settings, as opposed to Bayesian optimization where only the best solution is of interest, in this situation it provides clear evidence for this diminishing advantage.

This case study demonstrates the potential benefit of including human expertise and intuition to enhance the optimization of complex, real-world systems such as the simultaneous optimization of reactor geometry and operating conditions.

\subsection{Fed-batch Bioprocess Control}

For this case study, each participant is given the problem context described above, and the ability to select any number of initial solutions below 8.
They can view previous solutions, a visualization of current choices at each iteration, and choice information such as predicted mean objective. 

\begin{figure}[htb!]
    \centering
    \includegraphics[width=\textwidth]{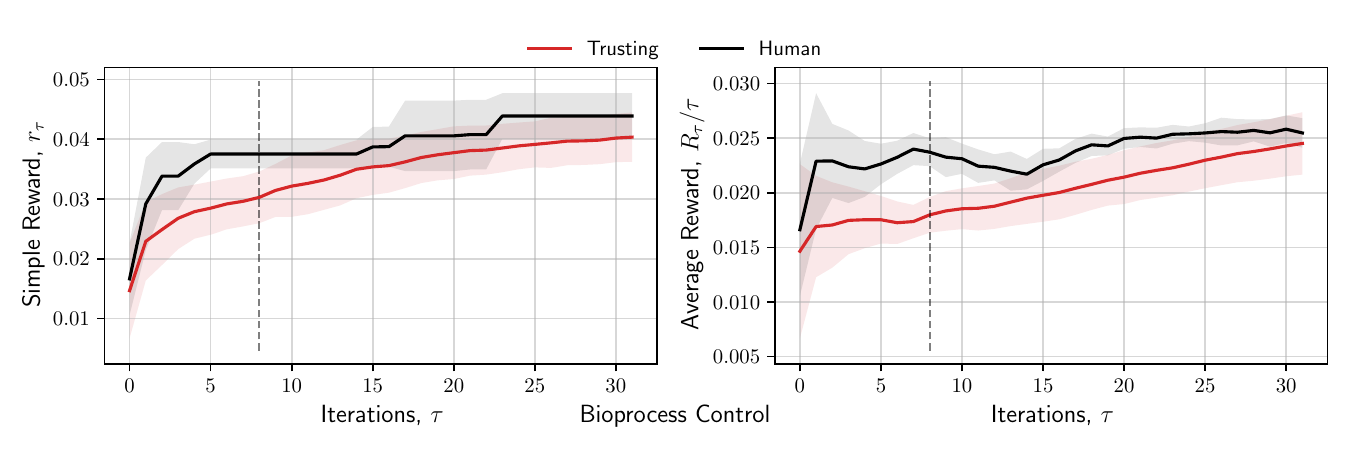}
    \caption{Simple and average reward presented alongside the reward obtained from standard Bayesian optimization. Filled sections represent one standard deviation and solid lines represent the mean reward.}
    \label{fig:bioprocess}
\end{figure}

Figure \ref{fig:bioprocess} demonstrates the average and standard deviation of reward, corresponding to the final product concentration, of the participants against standard Bayesian optimization.

On average, the inclusion of human-input throughout Bayesian optimization improves the convergence when compared to standard Bayesian optimization. 
We note that significant improvements are made during the initial sampling stage, where the participants had the ability to select initial solutions. 

We find that the initial solutions selected largely followed heuristics such as constant light or nitrogen feed rate, or ramps up or down. 
These heuristics prove more effective in this case study than a random control profile providing improved initial convergence. 
Overall, we demonstrate how the combination of the initial data generation, and continuous input contribute to improved performance.

\subsubsection{Qualitative Feedback}

When performing these experiments, an amount of qualitative feedback was generated from the participants as they were performing the task. In this subsection we present a number of these thoughts for discussion in groups.
\begin{center}
\textit{\textbf{Participant 1}: I wish I could say that I don't like \textbf{any} of the solutions.}\\
\textit{\textbf{Participant 3}: Can I say that I don't want to see that solution again?}
\end{center}

Expressing a negative sentiment for a solution was a common theme among participants. 
In the event that no solutions are desirable, our framework does enable the expert to simply choose their own (though in our benchmarking we choose not to allow this). 
Alternatively, the solution with the largest utility value can be selected. 
We choose not to incorporate the ability to express a negative choice in our approach as it may result in biasing the search space (through a weighted function or otherwise).
Future work may consider the inclusion of preference-based models as well as choice, however when taking this 'negative' viewpoint, care must be taken that globally optimal solutions are not erroneously disregarded. 
In comparison our approach can be considered broadly `optimistic' as no human bias is able to be introduced to the model itself, only at the stochastic decision making stage.

\begin{center}
\textit{\textbf{Participant 2}: I know this won't be a good solution, but I'll select it to stop it getting recommended.}\\
\textit{\textbf{Participant 3}: I will choose this one because I want to see more of that type of profile}
\end{center}

In these discussions, participants demonstrate non-myopic reasoning (that is, past one-step ahead).
Our approach enables this reasoning to be included within standard Bayesian optimization, however our acquisition function remains myopic avoiding the computational expense of a non-myopic acquisition function.
For a full treatment, a non-myopic acquisition function may be used.
Participant 2 raises an undesirable feature of our approach which, where undesirable solutions that a human chooses not to select are not evaluated, and therefore consistently proposed.
Non-myopic reasoning (selecting an undesirable solution) serves to break the deadlock in this case, but future work may be concerned with ensuring that solutions are selected that are different from previous iterations.

\section{Conclusions}

In this paper we present an approach for including expert-opinion continuously throughout Bayesian optimization, minimizing expert-time and effort.
Our approach relies on the proposal of a number of promising solutions from which the expert selects, these solutions are defined through the combination of high-throughput Bayesian optimization and multi-objective optimization. 
For a large class of problems our approach enables improvement on standard Bayesian optimization, with benefits maintained across high dimensional, noisy, and real world case studies. 
We demonstrate our approach using human participants on a real-world bioprocess optimization case study, as well as for the simultaneous design of coiled tube reactor operating conditions and geometry, achieving improved convergence over standard Bayesian optimization.

This work presents a step towards the inclusion of domain experts within Bayesian optimization, linking discrete decision making with existing state-of-the art computational methodologies for the optimization of engineering systems.

\section{Acknowledgments}

We would like to thank all the participants for their time in assisting us to benchmark our approach, A. Albay, B. Lewis, M. Bloor, A. Ahmed, M. de Carvalho Servia, V. Dehon. 
We would like to thank N. Basha for assistance in generating the original data used for the reactor geometry design case study.
Tom Savage would like to acknowledge the Imperial College President's Scholarship fund.

\appendix

\section{Optimization Benchmark Functions}

\subsection{Rastrigin}
\begin{align}
    f(\mathbf{x}) = 10d + \sum_{i=1}^d \left[ x_i^2 - 10 \cos(2 \pi x_i) \right] 
\end{align}

\subsection{Rosenbrock}

\begin{align}
    f(\mathbf{x}) = \sum_{i=1}^{d-1} \left[ 100 (x_{i+1} - x_i^2)^2 + (x_i - 1)^2 \right]
\end{align}

\subsection{Ackley}
\begin{align}
    f(\mathbf{x}) = -a \exp \left( -b \sqrt{\frac{1}{d} \sum_{i=1}^d x_i^2} \right) - \exp \left( \frac{1}{d} \sum_{i=1}^d \cos(c x_i) \right) + a + \exp(1)
\end{align}
where $a = 20$, $b = 0.2$, $c = 2 \pi$.

\subsection{Schewefel}

\begin{align}
    f(\mathbf{x}) = 418.9829d-\sum^d_{i=1}x_i\sin(\sqrt{x_i})
\end{align}

\section{Sampled Benchmark Functions}

Here we visualise sampled functions from a Gaussian process prior in order to provide further intuition into our results.

\begin{figure}[H]
    \centering
    \includegraphics[width=\textwidth]{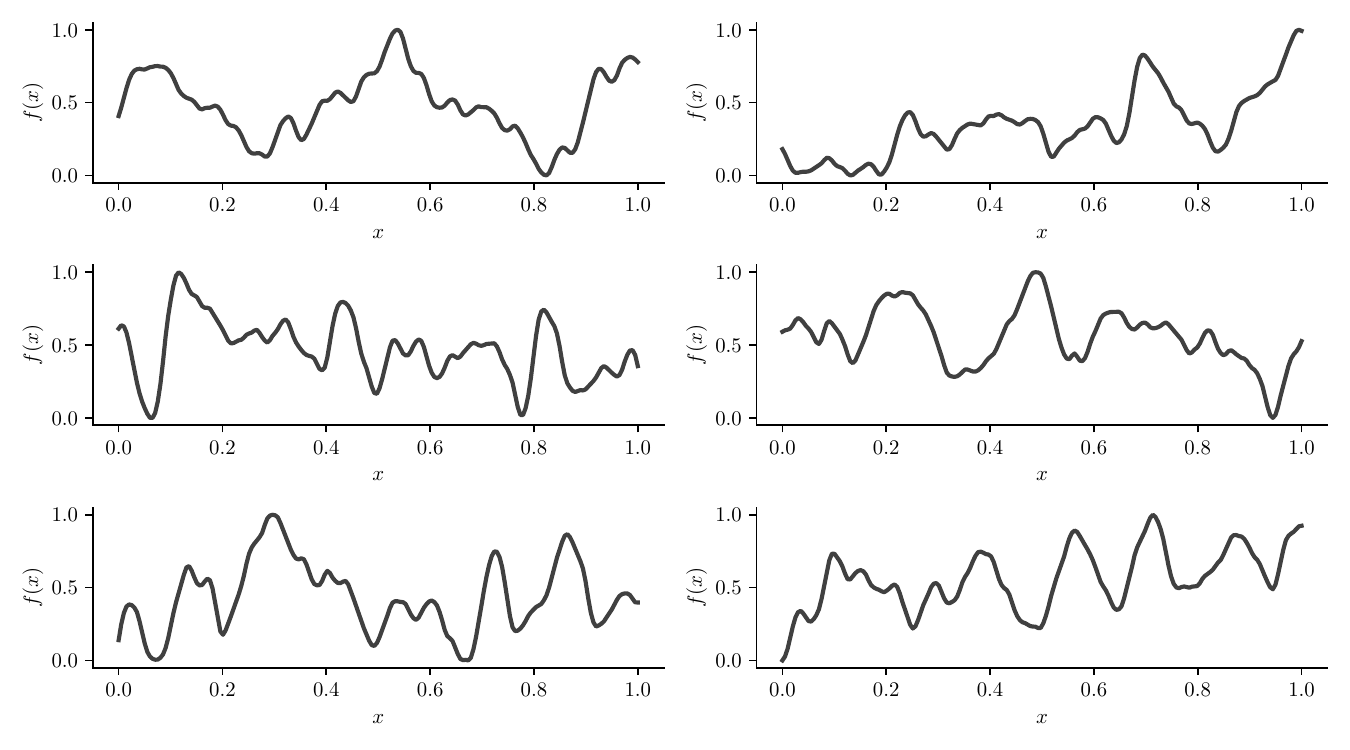}
    \caption{Functions sampled from a Gaussian process prior used for 1D benchmarking $f \sim \mathcal{GP}(\mu \equiv 0, K_M (d,\nu = 0.05))$ where $K_M$ is the Mat\'ern 5/2 kernel function.}\label{1DSampled}
\end{figure}

Figure \ref{1DSampled} demonstrates 1D functions sampled from a Gaussian process prior with Mat\'ern kernel.

\begin{figure}[H]
    \centering
    \includegraphics[width=\textwidth]{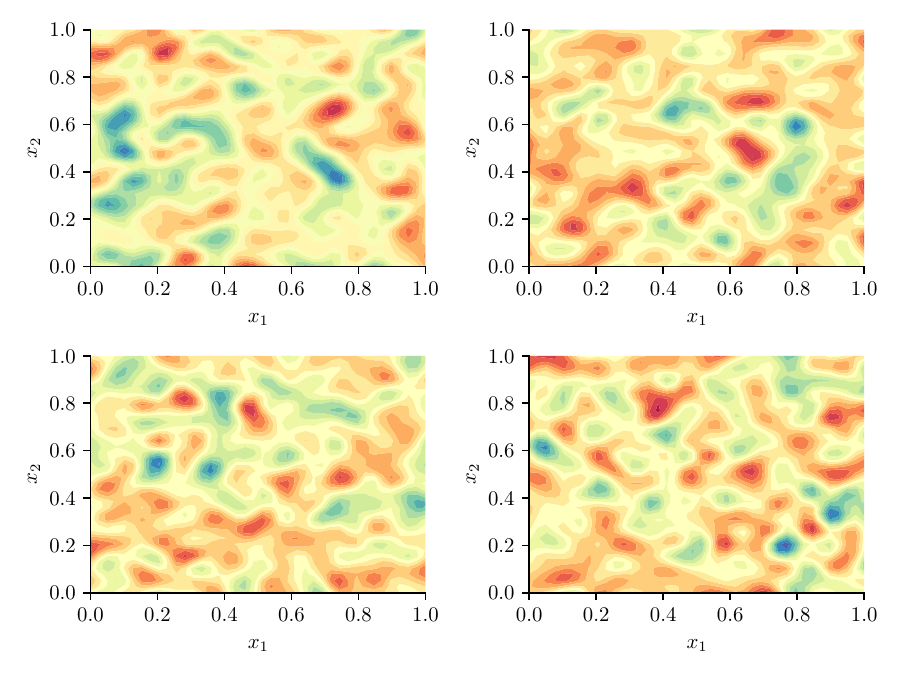}
    \caption{Functions sampled from a Gaussian process prior used for 2D benchmarking $f \sim \mathcal{GP}(\mu \equiv 0, K_M (d,\nu = 0.04))$ where $K_M$ is the Mat\'ern 5/2 kernel function.}\label{2DSampled}
\end{figure}

Figure \ref{2DSampled} demonstrates 2D functions sampled from a Gaussian process prior with Mat\'ern kernel.

In order to generate these functions, a Gaussian process prior was sampled at $300d$ locations throughout the unit bounds. 
These samples were then used to used within a noiseless Gaussian process which served as a callable proxy to the underlying sampled function. 
However, any approach that interpolates between points may be used. 
All code for the function generation can be located within the associated Github repository.

\section{Effect of acquisition function}

In addition to other results we present the effect of the specific acquisition function here. 
Whilst the acquisition function itself contains a number of assumptions, differences can result in improved performance \citep{logEI}. 

\begin{figure}[htb!]
    \centering
    \begin{subfigure}{\textwidth}
    \includegraphics[width=\textwidth]{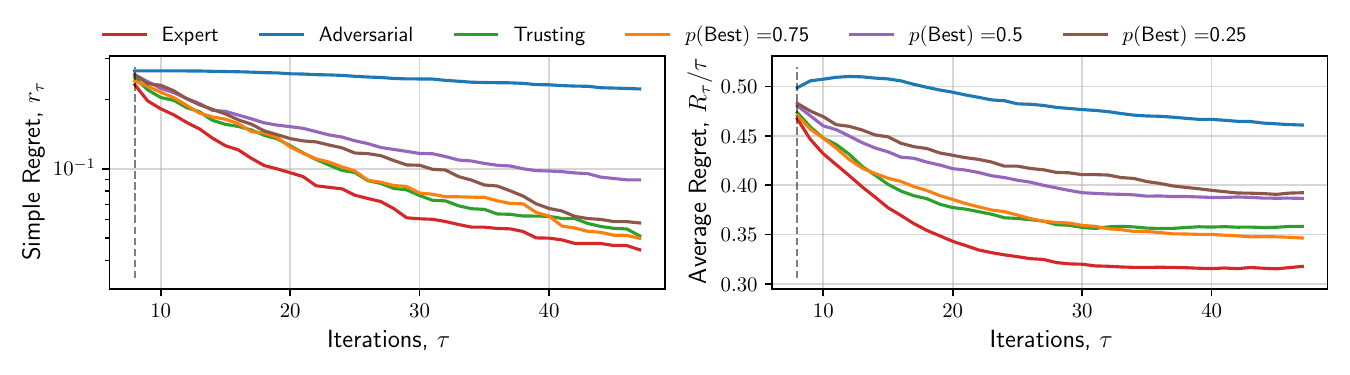}
    \caption{Expected Improvement.}
    \end{subfigure}
    \begin{subfigure}{\textwidth}
    \includegraphics[width=\textwidth]{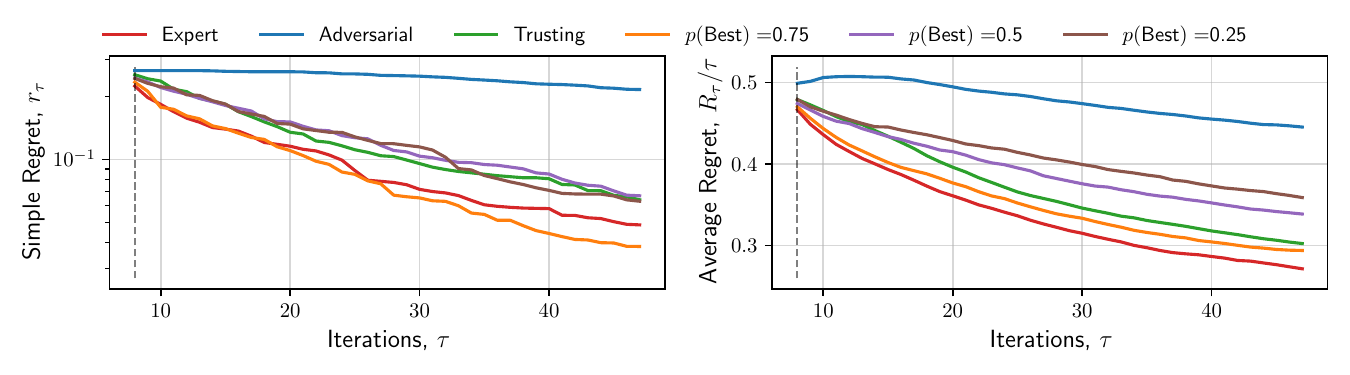}
    \caption{Upper Confidence Bound.}
    \end{subfigure}
    \caption{The effect of the acquisition function on the difference in convergence for sampled 2D functions.}
    \label{aq_diff}
\end{figure}

Figure \ref{aq_diff} demonstrates the performance of expected improvement and upper confidence bound for noiseless 2 dimensional functions sampled from a Gaussian process prior as previously described.

\begin{figure}[htb!]
    \centering
    \begin{subfigure}{\textwidth}
    \includegraphics[width=\textwidth]{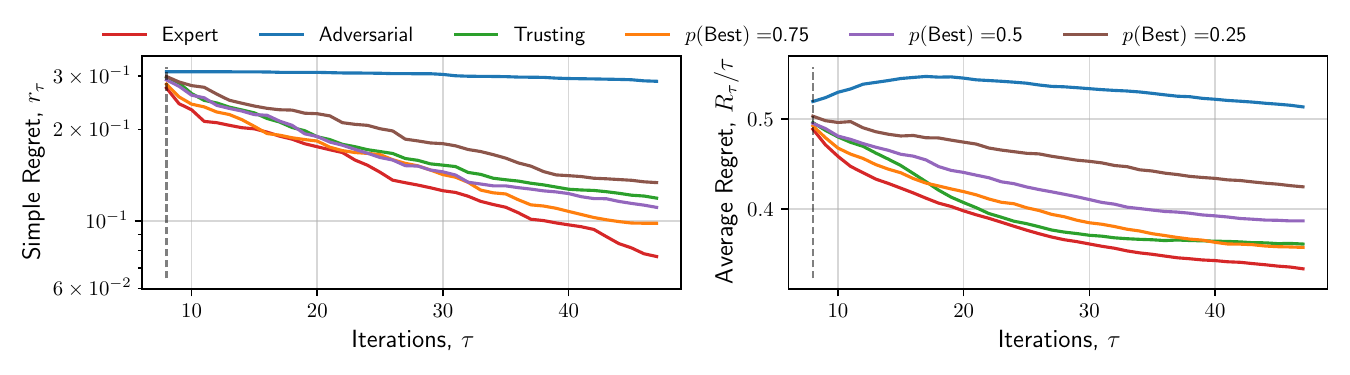}
    \caption{Expected Improvement.}
    \end{subfigure}
    \begin{subfigure}{\textwidth}
    \includegraphics[width=\textwidth]{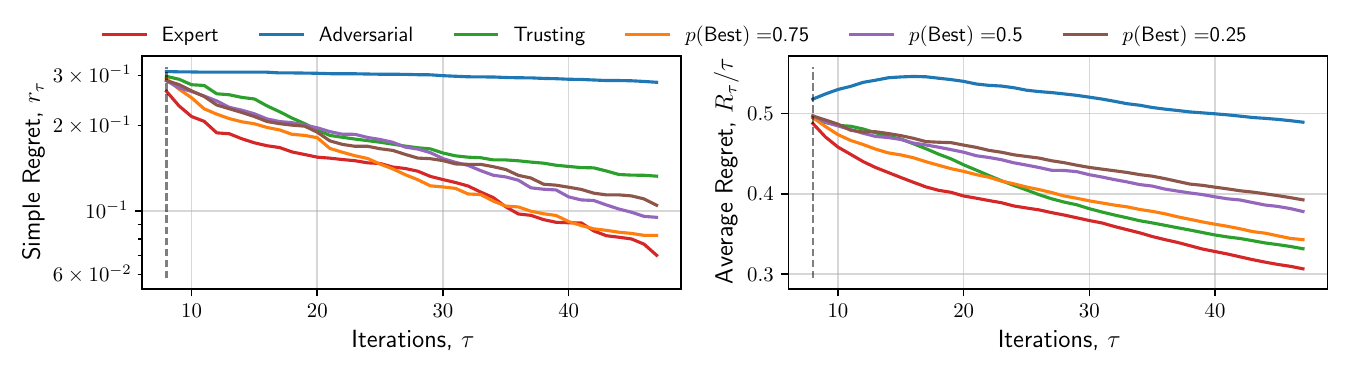}
    \caption{Upper Confidence Bound.}
    \end{subfigure}
    \caption{The effect of the acquisition function on the difference in convergence for sampled 3D functions.}
    \label{aq_diff_3}
\end{figure}

Figure \ref{aq_diff_3} demonstrates similar results across sampled 3D functions. 

Overall performance and trends across the two acquisition functions is largely similar.
Final regret values for the upper confidence bound function are slightly lower than expected improvement, potentially due to the more pronounced local optima than expected improvement. 
However, the trend is too small to draw concrete conclusions.

\newpage
\bibliographystyle{unsrtnat}
\bibliography{references}

\end{document}